\documentclass{ws-procs9x6}

\begin{document}

\title{Analyzing $\eta'$ Photoproduction Data on the Proton at
Energies of 1.5--2.3 GeV%
}

\author{K. Nakayama}

\address{Department of Physics and Astronomy,\\
University of Georgia,
Athens, GA 30602, USA\\
E-mail: nakayama@uga.edu}

\author{H. Haberzettl}

\address{Center for Nuclear Studies, Department of Physics,\\
The George Washington University,
Washington, DC 20052, USA \\
E-mail: helmut@gwu.edu}

\maketitle

\abstracts{
A combined analysis of the existing data on the reactions $\gamma p \to p
\eta^\prime $ and $pp \to pp\eta^\prime $, based on a relativistic meson
exchange model of hadronic interactions, is presented.}

\section{Introduction}

One of the primary interests in investigating $\eta^\prime$ meson-production
reactions is that they may be suited to extract information on nucleon
resonances, $N^*$, in the less-explored higher $N^*$ mass region. Current
knowledge of most of the nucleon resonances is mainly due to the study of $\pi
N$ scattering and/or pion photoproduction off the nucleon. Since the $\eta'$
meson is much heavier than a pion, $\eta'$ meson-production processes near
threshold necessarily sample a much higher resonance-mass region than the
corresponding pion production processes. Therefore, they are well-suited for
investigating high-mass resonances in low partial-wave states. Furthermore,
these reactions provide opportunities to study those resonances that couple
only weakly to pions, in particular, those referred to as ``missing
resonances''. 

Another special interest in $\eta ^{\prime }$ photoproduction is the
possibility to impose a more stringent constraint on its yet poorly known
coupling strength to the nucleon. This has attracted much attention in
connection with the so-called ``nucleon-spin crisis'' in polarized deep
inelastic lepton scattering\cite{EMC88} in that the $NN\eta^\prime$ coupling
constant can be related to the quark contribution to the ``spin'' of the
nucleon.\cite{Shore} Reaction processes where the $\eta^\prime$ meson is
produced directly off a nucleon may offer a unique opportunity to extract this
coupling constant.

\begin{figure}[t]
\centering
\includegraphics[width=.6\textwidth,angle=0,clip]{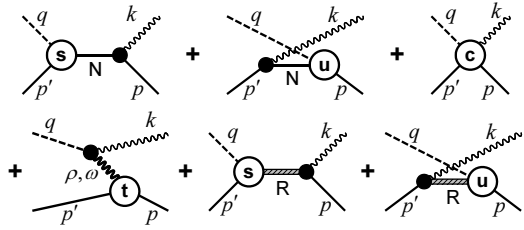}
\caption{\label{fig:1}%
Diagrams contributing to $\gamma p \to \eta' p$. The intermediate baryon states
are denoted \textsf{N} for the nucleon, and \textsf{R} for the nucleon
resonances. The total current is made gauge-invariant by an appropriate choice
of the contact current depicted in the top-right diagram. The nucleonic current
(nuc) referred to in the text corresponds to the top line of diagrams; the
mesonic current (mec) and resonance current contributions correspond,
respectively, to the leftmost diagram and the two diagrams on the right of the
bottom line of diagrams. }
\end{figure}

The major purpose of the present contribution is to present the photoproduction
part of a combined analysis of the $\gamma p \to p \eta^\prime$ and $pp
\rightarrow pp\eta^\prime$ reactions within a relativistic meson-exchange model
of hadronic interactions. The photoproduction reaction is described in the
tree-level approximation, where a phenomenological contact term is introduced
in order to guarantee the gauge-invariance of the full amplitude.\cite{NH1} The
latter consists of nucleonic, mesonic and (nucleon) resonance currents as
depicted in Fig.~\ref{fig:1}. The hadro-production reaction part of our
analysis as well as further details can be found in Ref.~\refcite{NH1}.

%


\section{Analysis of the SAPHIR data on $\gamma p \to p \eta^\prime$}
\label{sec:SAPHIR}

The objectives of analyzing the SAPHIR data\cite{SAPHIR} on $\eta^\prime$
photoproduction are:
\begin{itemize}
\item[1)]
To shed light on the conflicting conclusions of the existing model calculations
for these data. These contradictions are: (a) The origin of the shape of the
observed angular distribution. Zhao\cite{Zhao} has emphasized that this is due
to the interference among the resonance currents, while Chiang et
al.\cite{Chiang} have concluded that it is due to the interference between the
resonance and $t$-channel mesonic currents. Yet, Sibirtsev et
al.\cite{Sibirtsev} have claimed that the mesonic current is responsible for
the observed angular distribution. The latter authors use a $t$-dependent
exponential form factor in their mesonic current. (b) $t$-channel meson
exchange versus Regge trajectory. Chiang et al.\cite{Chiang} have emphasized
that the SAPHIR data can be described only if the Regge trajectory is used in
the $t$-channel mesonic current, while other authors\cite{Sibirtsev,Borasoy}
have used ordinary vector meson exchanges.

\item[2)]
Can we constrain the $NN\eta^\prime$ coupling constant from the photoproduction reaction?

\item[3)]
To which extent are we able to identify the nucleon resonances from the
differential cross section data, i.e., can, for example, the mass of the
resonance be pinned down from the existing cross-section data?

\item[4)]
Provide inputs for the $NN \to NN\eta^\prime$ reaction.
\end{itemize}

\begin{figure*}[t!]
\centering
\includegraphics[height=.812\textwidth,angle=-90,clip]{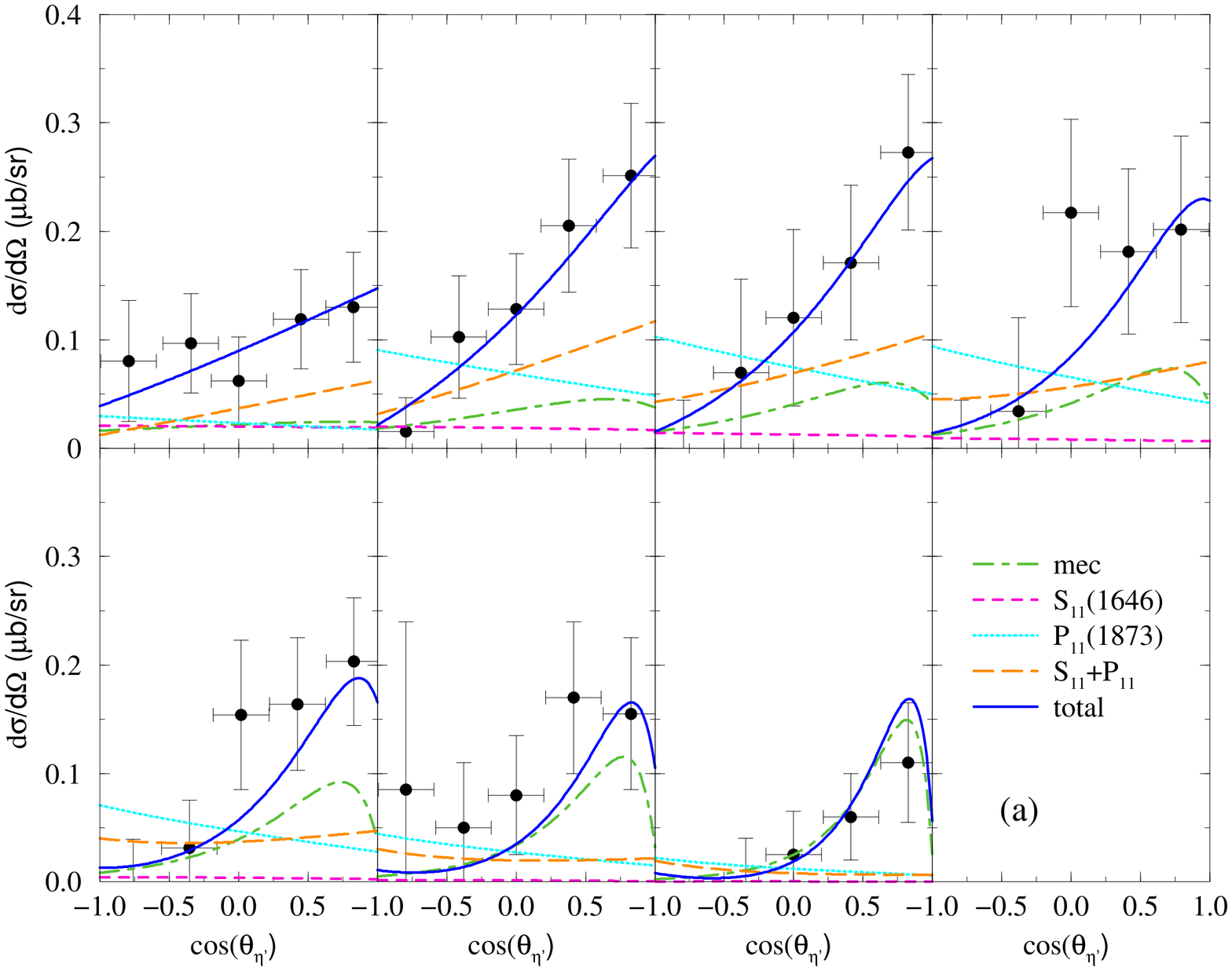}\\
\includegraphics[height=.812\textwidth,angle=-90,clip]{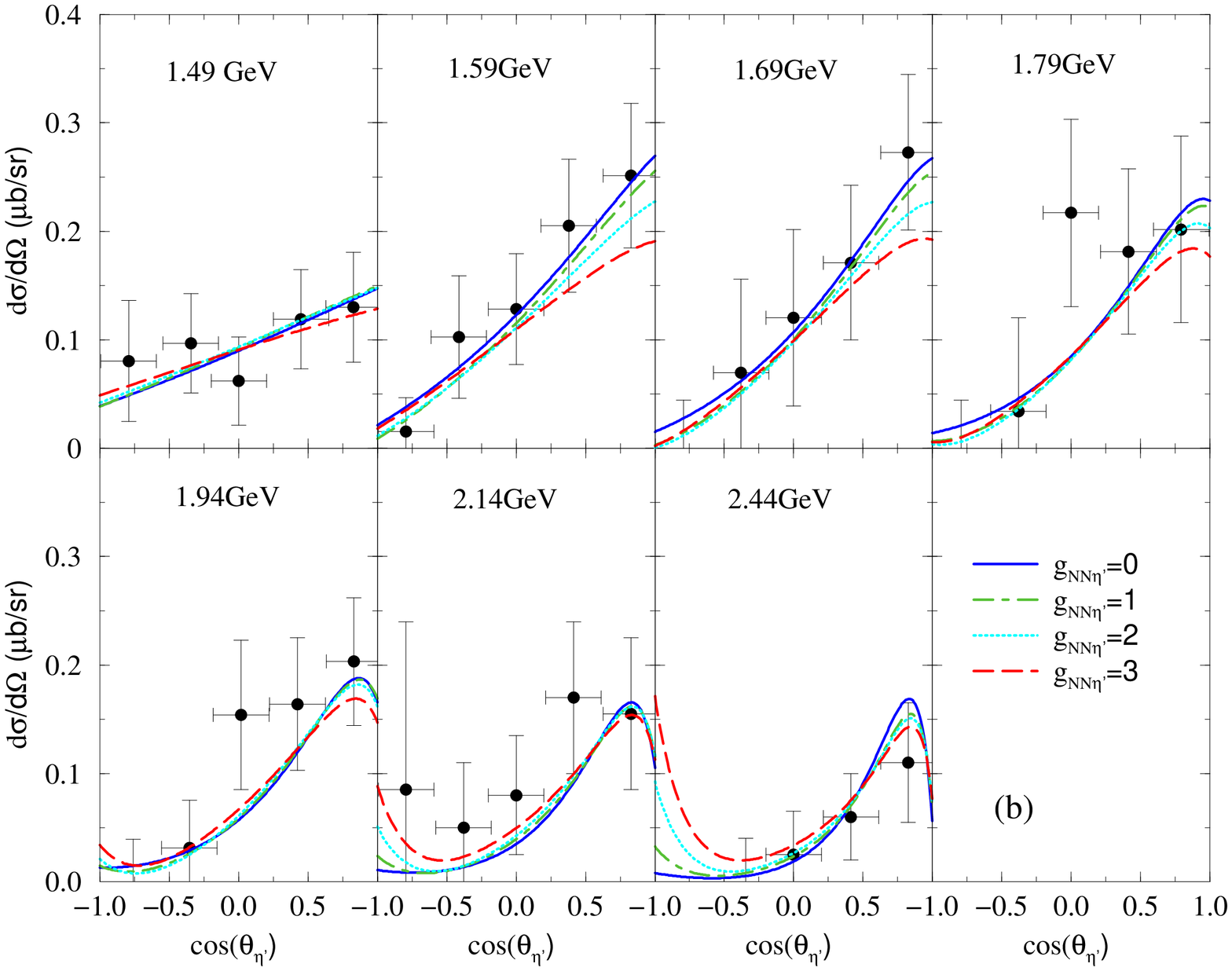}
\caption{\label{fig:3}%
Differential cross section for $\gamma p\to p \eta'$ according to the
mechanisms shown in Fig.~\ref{fig:1}a. Panel (a) includes the meson-exchange
current (mec), the $S_{11}$ and $P_{11}$ resonances. In (b), successively
stronger (as indicated by the values of the $g_{NN\eta'}$ coupling constant)
nucleonic current (nuc) contributions are added to the results shown in panel
(a). In each case, the model parameters are determined by best fits. The
meaning of the corresponding lines is indicated in the panels. The data are
from Ref.~\protect\refcite{SAPHIR}.}
\end{figure*}
%

Our results for the angular distribution at various energies are shown in
Fig.~\ref{fig:3}, together with the SAPHIR data.\cite{SAPHIR} First of all, the
mesonic current is responsible for the measured forward peaked angular shape at
higher energies. The $P_{11}$ resonance contribution is much larger than that
due to the $S_{11}$ resonance. Also, note the interference effects among
different currents; the $P_{11}$ resonance gives raise to a larger backward
angle cross sections, while the total resonance current, $S_{11}$ plus
$P_{11}$, yields a larger forward-angle cross sections. Adding the mesonic
current leads to a further enhancement of the forward cross sections. We
mention that the inclusion of the mesonic and $S_{11}$ resonance currents only
is not sufficient to describe the strong angular dependence at lower
energies.\cite{NH1} Also, the nucleonic current, through its interference with
the mesonic and $S_{11}$ resonance currents, makes the angular distribution
more pronounced,\cite{NH1} but not as pronounced as the addition of the
$P_{11}$ resonance shown in Fig.~\ref{fig:3}a. It is clear, therefore, that the
observed angular distribution is a result of the rather non-trivial
interference among different currents.

%

Figure~\ref{fig:3}b displays our results for various values of the
$NN\eta^\prime$ coupling constant, $g_{NN\eta^\prime}$, in the nucleonic
current. Here, we also include the mesonic and the $S_{11}$ and $P_{11}$
resonance currents. Note that the nucleonic current becomes pronounced at
backward angles as the energy increases which is due to the $u$-channel
diagram. We see that $g_{NN\eta^\prime}$ cannot be much larger than 3. Naively,
we would expect that more accurate data at higher energies will enable us to
reduce this upper limit. See, however, the analysis of the high-precision CLAS
data in Sec.~\ref{sec:CLAS}.

Next, we address the issue of the ordinary meson exchange versus Regge
trajectory in the $t$-channel mesonic current. The mesonic current based on the
ordinary vector-meson exchange contains an extra form factor at the
electromagnetic vertex, while that based on the Regge trajectory contains no
such form factor. We verified that both models describe the data equally well,
although there are some differences in detail. However, one important point to
be noted here is that the resulting resonance parameters are quite different.
This is quite disturbing, for it reveals a clear model dependence in the
extracted resonance parameters. Further investigation of this important issue
is required in future works.

We have also verified the sensitivity/insensitivity of the differential cross section data
to the mass value of the nucleon resonance. In particular, we found that the results with
the $S_{11}$ mass values which differ by about 100 MeV  are hardly distinguishable from each
other in the differential cross sections. This gives a rough idea about the uncertainty one
should expect on the extracted resonance mass values based only on the differential cross
section data.

\section{Analysis of the (preliminary) CLAS data on $\gamma p \to p \eta^\prime$ reaction}
\label{sec:CLAS}

\begin{figure}[t!]
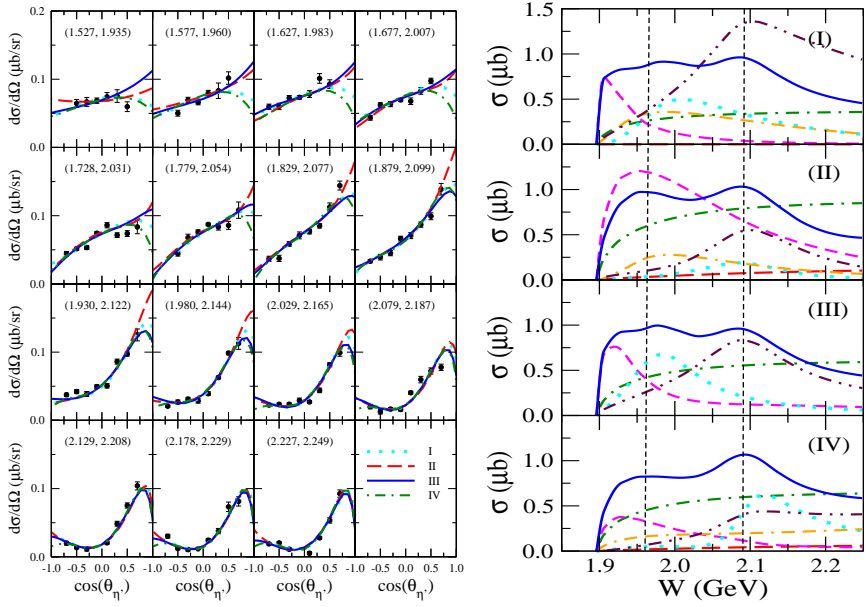

\includegraphics[width=2.4in,angle=0,clip]{dxscRall.eps}
\hfill
\includegraphics[height=3.15in,width=2in,angle=0,clip]{txscRall.eps}
\caption{\label{fig:7}%
Left figure (a): Same as Fig.~\ref{fig:3} for the CLAS data.\protect\cite{CLAS}
The curves correspond to different fit results which yield comparable $\chi^2$
values. The numbers ($T_\gamma , W$) in parentheses are the incident photon
energy $T_\gamma$ and the corresponding $s$-channel energy $W=\sqrt{s}$,
respectively, in GeV. ~~Right figure (b): Total cross section for $\gamma p\to
p \eta'$ as a function of $W$. As indicated in the legend, the panels
correspond to the fit results shown in the left figure. The overall total cross
sections (solid lines) are broken down according to their dynamical
contributions. The dash-dotted curves correspond to the mesonic current
contribution; the dashed curves to the $S_{11}$ resonance current and the
dotted curves to the $P_{11}$ resonance. The dot--double-dashed curves
correspond to the $P_{13}$ resonance current while the dash--double-dotted
curves show the $D_{13}$ resonance contribution. The nucleonic current
contribution (long-dashed curves) are negligible and cannot be seen on the
present scale. The two dashed vertical lines are placed to guide the eye
through the two bump positions in all panels. }
\end{figure}

Our model results for the (preliminary) CLAS data on $\eta^\prime$
photoproduction\cite{CLAS} are shown in Fig.~\ref{fig:7}a. First of all,
compared to the SAPHIR data\cite{SAPHIR} analyzed in Sec.~\ref{sec:SAPHIR}, the
new CLAS data are much more accurate and, as such, may reveal features that
were not seen in the analysis of the SAPHIR data. In Fig.~\ref{fig:7}a,
different curves correspond to different sets of fit parameters which yield
comparable $\chi^2$ values. Unlike the case of the SAPHIR data, here one
requires not only the spin-1/2 resonances, but also the spin-3/2 resonances in
order to reproduce the data. We found that the required spin-1/2 and -3/2
resonances are consistent with those quoted by the PDG.\cite{PDG}

%

Although the different parameter sets yield practically the same differential
cross sections (except for very forward and backward angles where no data
exist), the corresponding dynamical content is very different from each other
over the entire angular range. This shows, in particular, that cross sections
alone are unable to fix the resonance parameters unambiguously, and that more
exclusive observables, such as the beam and target asymmetries, are necessary
in order to extract information on nucleon resonances.

Our predictions for the total cross section, as displayed in Fig.~\ref{fig:7}b,
have been obtained by integrating the differential cross section results of
Fig.~\ref{fig:7}a. A common feature present in all of these results is the bump
structure around $W=2.09$ GeV. If this is confirmed, the $D_{13}(2080)$ [and
possibly $P_{11}(2100)$] resonance is likely to be responsible for the
structure. The PDG\cite{PDG} quotes $D_{13}(2080)$ and $P_{13}(2100)$ as two
and one star resonances, respectively. Another feature we see in
Fig.~\ref{fig:7}b is the sharp rise of the total cross section near threshold
which is caused by the $S_{11}$ resonance.

Contrary to the expectation in Sec.~\ref{sec:SAPHIR}, the $NN\eta^\prime$
coupling constant cannot be determined even with the high-precision CLAS data.
The reason is that even at higher energies, the resonance contribution may be
large, especially that of the $D_{13}$ resonance. In fact, two of the results
shown in Fig.~\ref{fig:7}b correspond to practically vanishing
$g_{NN\eta^\prime}$. However, we are able to give a more stringent upper limit
of $g_{NN\eta^\prime}<2$. In order to pin down further this coupling constant,
one needs to either have more exclusive data than the cross section or go
beyond the resonance energy region.

\section{Summary}
\label{sec:summary}

The study of $\eta^\prime$ production processes is still in its early stage of
development. Our study reveals that in order to extract relevant physics, one
needs more exclusive data than the cross sections. In addition, measurements of
$\eta^\prime$ production using the neutron/deuteron target are also required.
On the other hand, our theoretical model also
needs to be improved; in particular, coupled channel effects should be
investigated. In this connection, unfortunately, there is no realistic model for
the $N\eta^\prime$ final state interaction available at present.

\section*{Acknowledgments}
This work is partly supported by the Forschungszentrum J\"ulich, COSY Grant
No.\ 41445282\;(COSY-58).

\end{document}